\documentclass[useAMS]{mn2e}
\usepackage{amssymb,graphicx,longtable,lscape}
\usepackage{wasysym}

\def\msol{\hbox{$\rm\thinspace M_{\odot}\thinspace$}} 
\def\sol{\hbox{$\rm\thinspace _{\odot}\thinspace$}} 
\def\etal{{\it et al.\thinspace}}

\newcommand{\be}{\begin{equation}}
\newcommand{\ba}{\begin{eqnarray}}
\newcommand{\ee}{\end{equation}}
\newcommand{\ea}{\end{eqnarray}}

\title[On Type Ia Supernovae From The Collisions of Two White Dwarfs]{On Type Ia Supernovae From The Collisions of Two White Dwarfs}
\author[Cody Raskin, F.X. Timmes, Evan Scannapieco, Steven Diehl, \& Chris Fryer]
{Cody Raskin$^{1}$\thanks{E-mail: cody.raskin@asu.edu}, F.X. Timmes$^{1,3}$, Evan Scannapieco$^{1}$, Steven Diehl$^{2}$, \& Chris Fryer$^{2}$\\
$^{1}$School of Earth and Space Exploration, Arizona State University, Tempe, AZ\\
$^{2}$Los Alamos National Laboratories,  Los Alamos, NM\\
$^{3}$Joint Institute for Nuclear Astrophysics, Notre Dame, IN}

\begin{document}

\date{Accepted 2009 August 14}


\maketitle

\begin{abstract}
We explore collisions between two white dwarfs as a pathway for making
Type Ia Supernovae (SNIa).  White dwarf number densities in globular
clusters allow $10-100$ redshift $\lesssim1$ collisions per year, and
observations by (Chomiuk \etal 2008) of globular clusters in the
nearby S0 galaxy NGC 7457 have detected what is likely to be a SNIa
remnant.  We carry out simulations of the collision between two
0.6\msol white dwarfs at various impact parameters and mass
resolutions. For impact parameters less than half the radius of the
white dwarf, we find such collisions produce $\approx$ 0.4 M$_{\odot}$
of $^{56}$Ni, making such events potential candidates for
underluminous SNIa or a new class of transients between Novae and
SNIa.
\end{abstract}

\begin{keywords}
hydrodynamics -- nuclear reactions, nucleosynthesis, abundances -- (stars:) white dwarfs -- (stars:) supernovae: general.
\end{keywords}

\section{Introduction}

Type Ia supernovae (henceforth SNIa) play a key role in astrophysics
as premier distance indicators for cosmology
(Phillips 1993; Riess \etal 1998; Perlmutter \etal 1999), as direct
probes of low-mass star formation rates at cosmological distances
(Scannapieco \etal 2005; Mannucci \etal 2006; Maoz 2008)
and as significant contributors to iron-group elements in the cosmos
(Wheeler \etal 1989; Timmes \etal 1995; Feltzing \etal 2001; Strigari 2006).
Our current understanding is that there are two major progenitor
systems for these events.  The first possibility, the single-degenerate scenario,
consists of a carbon-oxygen white dwarf in a binary system evolving to
the stage of central ignition by mass overflow from a low-mass stellar
companion (Whelan \& Iben 1973; Nomoto 1982; Hillebrandt \& Niemeyer 2000).
The second possibility, the double-degenerate scenario, consists of
the merger of two white dwarfs in a binary system
(Iben \& Tutukov 1984; Webbink 1984; Yoon \etal 2007).  It is unknown 
at what relative
frequency both of these channels operate (Livio 2000; Maoz 2008).

Collisions between two white dwarfs, are likely to happen less
frequently than binary mergers. 
However, as discussed in Timmes (2009) and Rosswog \etal (2009), they will occur in 
globular clusters where the stellar densities are extremely high. 
 For a typical globular
cluster velocity dispersion of $\approx$ 5-10 km s$^{-1}$, and a white
dwarf escape velocity of $\approx$ 4000 km s$^{-1}$, the physical
cross-section of equal mass white dwarfs is enhanced by a factor of $4
\times 10^4 - 2 \times 10^5$ by gravitational focusing.  For an average globular cluster mass 
of $10^6$\msol (Brodie \& Strader 2006),
a Salpeter IMF (Salpeter 1955), and a globular cluster core radius
of 1.5 pc (Peterson \& King 1975), we conservatively estimate an average
white dwarf number density in globulars to be $\approx10^{4}$ pc$^{-3}$, 
and estimating that about $1/1000$ of the
average $5\times 10^8$ $M_\odot$ per Mpc$^{-3}$ stellar mass density
in the universe is in globular clusters (see Pfahl \etal 2009 and references therein) the 
overall rate of such collisions is about $5-20 \times 10^{-10}$ per comoving Mpc$^3$ per year.  
This corresponds to $10-100$, $z \lesssim 1$ collisions per year.  
Note that this calculation depends in detail on the minimum impact parameter 
that leads to explosive nuclear burning, as studied in detail below.
Our estimate is significantly lower than a recently submitted paper (Rosswog
\etal 2009), primarily due to the assumed core radius of globulars clusters.

Beyond $\approx$ 50 Mpc, current surveys are not sufficiently accurate
to distinguish between supernovae in globular clusters from those that
arise from the galaxy field stars in front or behind globular clusters
(Pfahl \etal 2009).  Furthermore, observations of globular clusters
in the nearby S0 galaxy NGC 7457 have detected what is likely to be a
remnant of a SNIa (Chomiuk \etal 2008). Taken together, these
estimates and observations suggest that white dwarf collisions are
likely to appear in current and future SNIa samples.  As
double-degenerate supernovae from collisions may not fit the standard
templates, SN surveys may have to consider possible contamination
from double-degenerate collisions that masquerade as traditional SNIa.

In this paper we begin to examine collisions between two white dwarfs as a
pathway for producing SNIa.  In \S\ref{sec:simulations} we describe
our simulations of the impact between two 0.6 M$_{\odot}$ white
dwarfs, and in \S\ref{sec:results} we discuss the resulting
hydrodynamics, thermodynamics, and nucleosynthesis.  In
\S\ref{sec:discussion}, we speculate on other mass pairs and impact
parameters, and discuss the implication of our results.

\section{SNSPH Simulations}
\label{sec:simulations}

All of our simulations are conducted with a 3D smooth particle
hydrodynamics code, SNSPH (Fryer \etal 2006).  In collision scenarios
with a non-zero impact parameter, angular momentum plays a critical
role in the final outcome, and Lagrangian particle methods generally
conserve angular momentum better than Eulerian grid methods. We added
a 13 isotope $\alpha$-chain nuclear reaction network (Timmes 1999;
Timmes \etal 2000; Fryxell \etal 2000) and a Helmholtz free energy
based stellar equation of state (Timmes \& Arnett 1999; Timmes \&
Swesty 2000) to SNSPH to construct the initial white dwarf
models and run our collision models.

Our initial 0.6\msol white dwarf is composed of equal parts $^{12}$C
and $^{16}$O and is created using the Weighted Voronoi Tessellations
method (Diehl \& Statler 2006), which arranges particles in the
configuration corresponding to the lowest energy state 
that is consistent with a given equation of state.  The resulting
white dwarfs have only minor temperature variations around
1$\times$10$^7$ K. Any evolution the white dwarfs experience before
the collision, then, is due only to their mutual gravitational
interaction.  We use a simple 2-body solver to calculate the relative
velocities of the two white dwarfs immediately before a collision for
a given impact parameter and an initial relative velocity of 10 km
s$^{-1}$, which corresponds to the typical virial velocity of a
globular cluster.  The velocities calculated by our 2-body solver then
serve as the initial conditions for our collision simulations.

Various angles of incidence for the collision were chosen.  In Table
1, the initial impact parameter, $b$, is the vertical separation of
the centers of the white dwarfs in terms of R\sol at a horizontal
separation of $\infty$. Due to gravitational focusing, the final
impact parameter, $d$, the vertical separation at the moment of
collision, is significantly smaller, and is given in Table 1 in terms
of the radius of a 0.6\msol white dwarf, $\approx 0.01$R\sol.
The number of particles $N$ in each model is also given in Table 1,
i.e., there are $N/2$ particles per white dwarf.

\begin{table}
\centering
\caption{Simulation impact parameters, $b$, $d$, and total particle count, $N$.}
\begin{tabular}{ c | c c | c }
\hline\hline
Scenario & $b$ [R\sol] & $d$ [R$_{wd}$] & $N$ [x1000] \\
\hline
1 & 0 & 0 & 800 \\
2 & 0.9 & 0.5 & 800 \\
3 & 1.7 & 0.9 & 200 \\
\hline
\end{tabular}
\label{table:cases}
\end{table}

\section{Results}
\label{sec:results}

\subsection{Scenario 1}

\begin{figure}
\centering
\includegraphics[width=7cm,height=10.5cm]{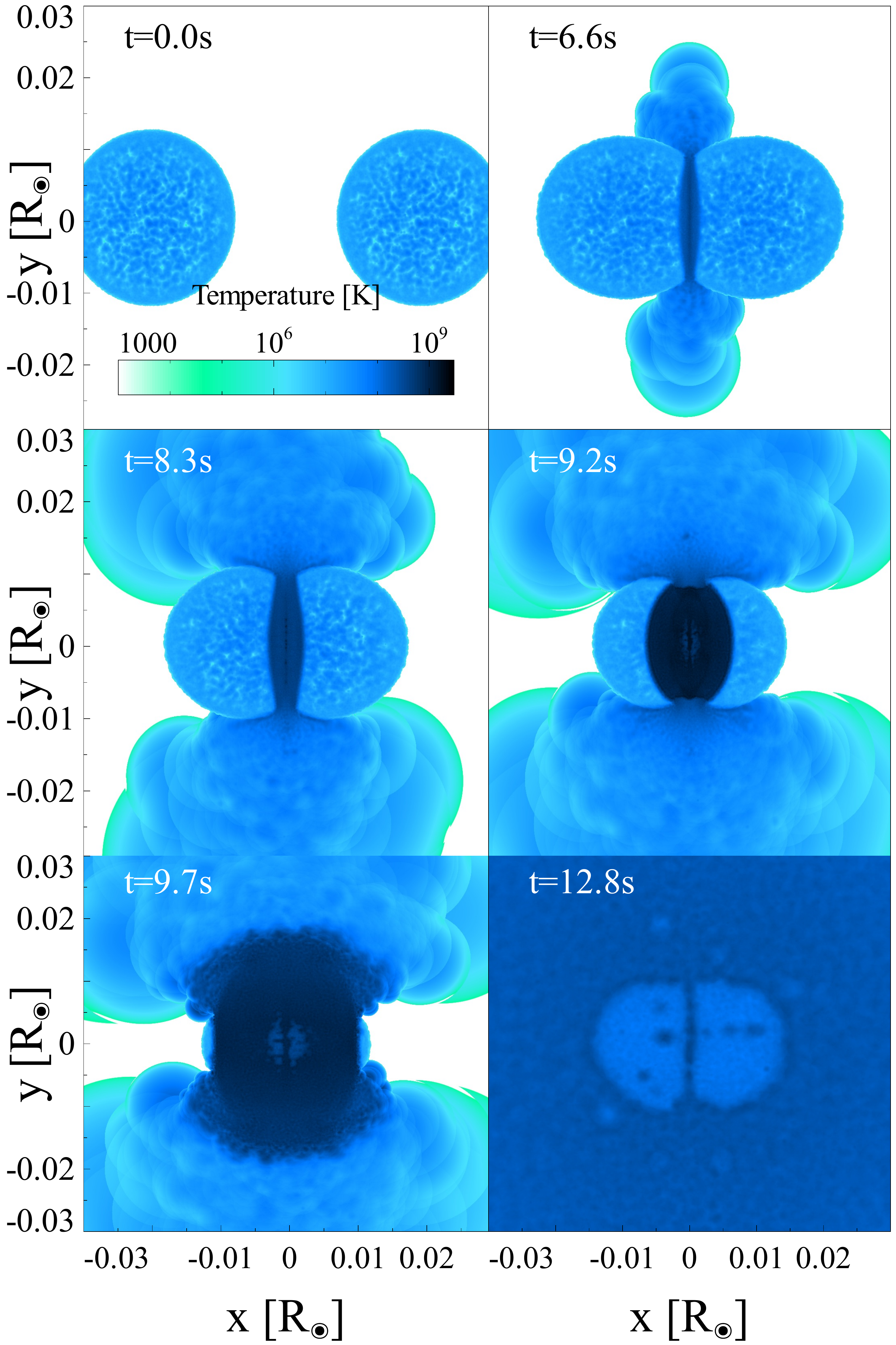}
\caption{Planar slices through the 3D calculation of the evolution of 
head-on white dwarf collision scenario 1, given in Table 1. The features 
of each panel are described in detail in \S 3.1. The sequence displayed, 
after the white dwarfs have collided in panel 2, takes place over a period 
of $\approx$6 seconds.}
\label{fig:headon}
\end{figure}

\begin{figure}
\centering
\includegraphics[width=7cm,height=7cm]{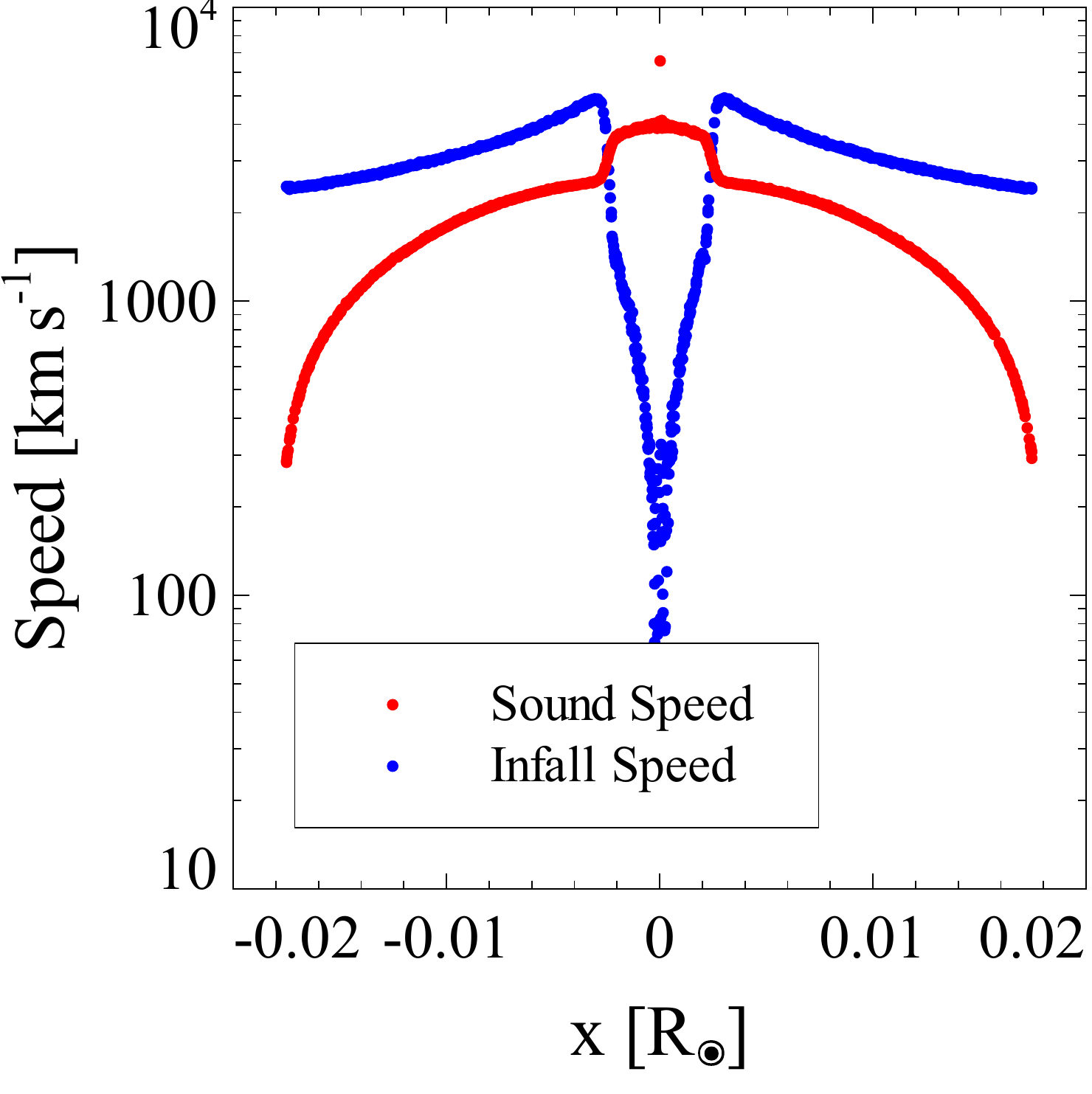}
\caption{The sound speed and infall speed of particles lying on the $x$-axis
 in scenario 1. This snapshot corresponds to panel 3 in Figure \ref{fig:headon}.
  The material falling into the shocked region is moving at or above the sound
   speed for this region, causing the shock to stall.}
\label{fig:velocities}
\end{figure}

Snapshots of planar slices from the head-on collision case are given
in Figure \ref{fig:headon}.  
This extremely unlikely, but physically instructive scenario
was presented in Timmes (2009) and has also recently been
explored in Rosswog \etal (2009).
As the white dwarfs approach each other,
their velocities increase to the escape speed, $\approx$4000 km
s$^{-1}$ (first panel).  When the two white dwarfs first make contact
(second panel), shock waves develop that attempt to travel outward
along the $x$-axis at roughly the sound speed, $\approx$3000 km
s$^{-1}$ (also see Figure \ref{fig:velocities}).  The near equality of
these two characteristic speeds means the shock waves stall as
material falls through them (Fig.\ \ref{fig:headon} third panel).  The
region between the two fronts achieves a nearly constant temperature
($T \sim 10^9$ K) and density ($\rho \sim 10^6$ g cm$^{-3}$) with a
non-explosive, mild rate of nuclear burning. This shocked region
expands very slowly as more material piles into it.  
In this regime the nuclear energy generation rate 
scales as $\dot \epsilon \approx \rho^2 \ T^{27}$.

When the high density ($\approx 4\times10^6$ g cm$^{-3}$) central core
of each white dwarf encounters the leading edge of the shock, it
raises the energy generation rate of enough material above the
threshold needed to trigger a detonation (see Gamezo \etal 1999 and references therein).  Two curved detonation fronts with
$T\sim10^{10}$ K form at the interfaces of the shocked and unshocked
regions and begin to propagate (fourth panel in Figure
\ref{fig:headon}) which releases enough energy to unbind the merged
system (fifth panel). The entire system then undergoes homologous
expansion (sixth panel), leaving a small amount of unburned CO in the
central regions surrounded by a layer of $^{56}$Ni and other
iron-group elements, a layer of Si-group elements, and then an outer
region of unburned CO.  The final tally, with 1.2\msol of CO entering
the system, is 0.34\msol of $^{56}$Ni, 0.33\msol of $^{28}$Si, and
0.15\msol of unburned CO with the remaining mass distributed among
other heavy elements (also see Figure \ref{fig:convergence}). Such an
event, whose total mass is sub-Chandrasekhar, would be a potential
candidate for either an underluminous SNIa or a new class of
transients between Novae and SNIa.

\subsection{Scenario 2}

\begin{figure}
\centering
\includegraphics[width=7cm,height=7cm]{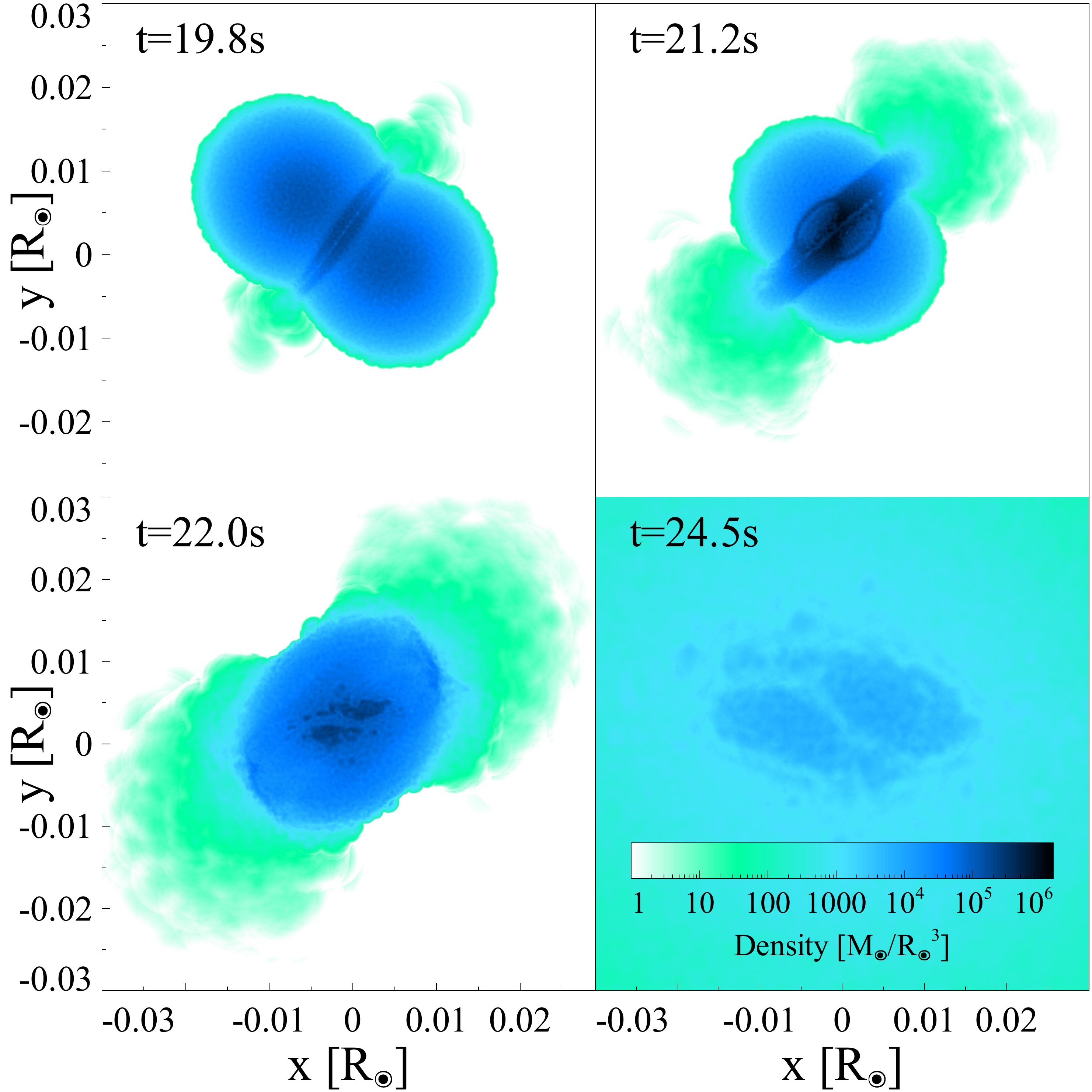}
\caption{Planar slices through the 3D calculation of the evolution of the $d = 0.5 {\rm R}_{wd}$
white dwarf collision scenario 2, given in Table 1. The features of each panel are described in detail in \S 3.2.}
\label{fig:grazing}
\end{figure}

In scenario 2, we use an impact parameter of $b=0.9$R\sol in order to
bring the stars into contact with a final impact parameter of
$d=0.5$R$_{wd}$ or $d\approx0.005$R\sol. As in scenario 1, the shock
front stalls as the inward falling material is traveling at a
comparable speed (first panel of Figure \ref{fig:grazing}). Once
again, when the cores of the white dwarfs enter the shocked region,
two detonation fronts break out and unbind the system (second
panel). However, in this case, the additional torque applied to the
shocked region as the stars continue to move past each other results
in off-axis and off-center detonation fronts that occur at earlier
times than in scenario 1 (second and third panels).
In this case, 0.30\msol of $^{56}$Ni, 0.37\msol of $^{28}$Si are
created, leaving 0.25\msol of unburned CO with the remaining mass
distributed among other heavy elements.

\subsection{Scenario 3}

In scenario 3, we increase the impact parameter to $b=1.7$R\sol,
resulting in a pre-collision impact parameter of $d=0.9$R$_{wd}$ or
$d\approx0.009$R\sol. This ensures a purely grazing incident, which is
likely to be the the most frequent impact scenario as a wide range of
impact parameters will result in a similar, grazing collision when
tidal effects transfer angular momentum from passing impactors.
Absent the violent shock experienced in scenarios 1 and 2, there is
negligible nuclear burning in a purely edge-on impact. Less than
$10^{-3}$\msol of $^{56}$Ni are produced by the initial
interaction. The kinematics of the collision causes both stars to
become unbound, forming a rotating disk of white dwarf debris, which
will eventually collapse into a single compact object as it cools.


\subsection{Convergence Studies}

\begin{figure}
\centering
\includegraphics[width=7cm,height=7cm]{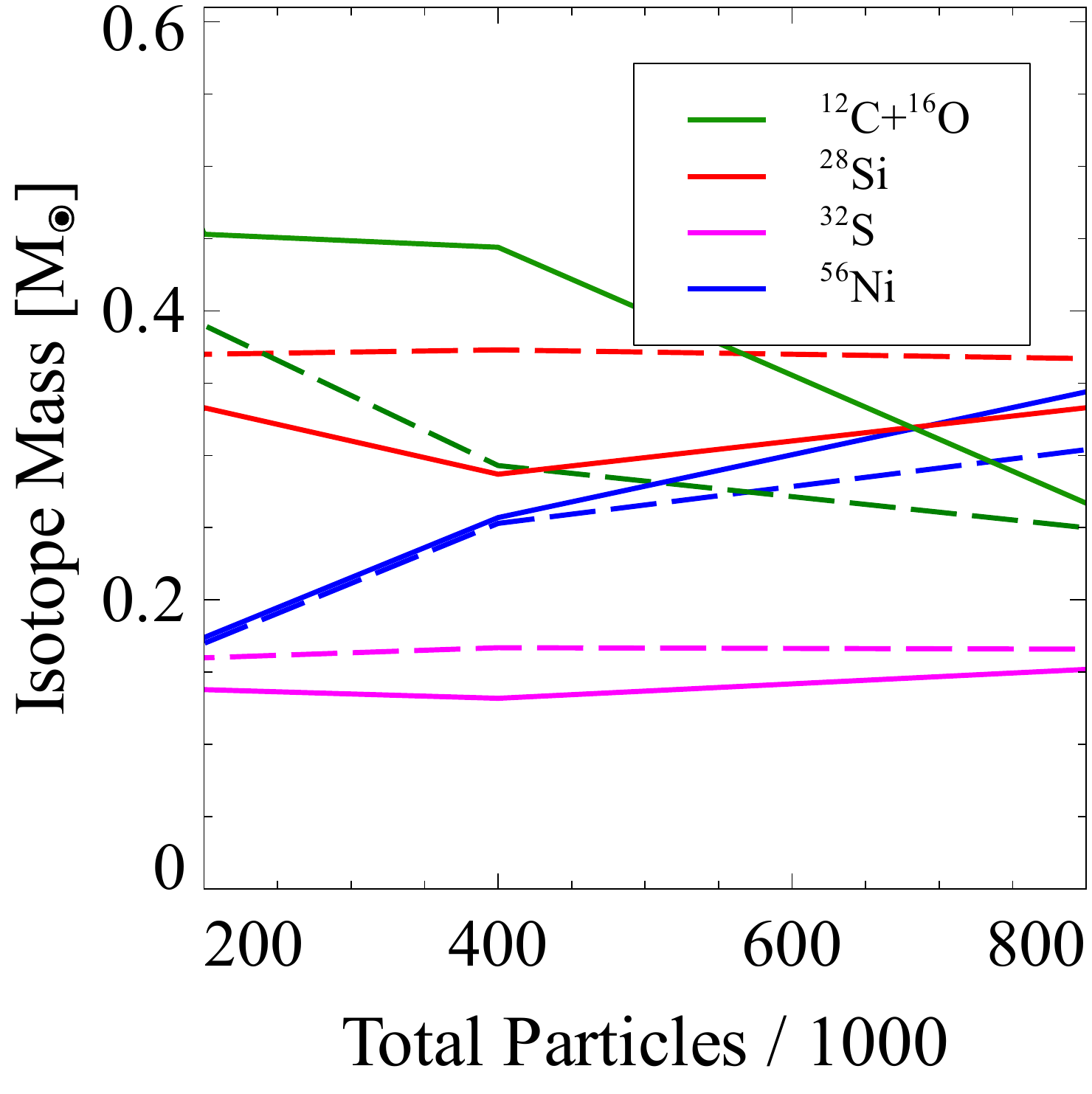}
\caption{Isotope yields vs.\ total particle count for scenarios 1 and 2. The solid lines are the isotope yields of the head-on simulation, while the dashed lines are those of the $d=0.5$R$_{wd}$ simulation.}
\label{fig:convergence}
\end{figure}

A known weakness of smooth particle hydrodynamic simulations is
resolving shocks with a finite number of particles.  For this reason,
we carried out a convergence study of the the $d=0$ (scenario 1) and
$d=0.5$R$_{wd}$ (scenario 2) cases in order to place upper limits on
the isotope yields.  Figure \ref{fig:convergence} shows that with
increasing particle count, more of the CO is converted into $^{56}$Ni
and $^{28}$Si, while the abundance of $^{32}$S remains relatively
constant. While we did not achieve an absolute convergence in our
simulations, we can reasonably extrapolate an upper limit of 0.4\msol
of $^{56}$Ni for the head-on merger scenario and 0.35\msol in the
$d=0.5$R$_{wd}$ scenario, where 1.2\msol of CO enters the system.
In low impact-parameter collision scenarios involving larger masses for 
the constituent white dwarfs, it is very likely that this yield would increase.

Rosswog \etal (2009) find 0.32 M\sol of $^{56}$Ni for the head-on
collision of two 0.6 M\sol white dwarfs with an SPH resolution of
2$\times$10$^5$ particles. This is consistent with our highest
resolution SNSPH model with 8$\times$10$^5$ particles, but is
inconsistent with our lower resolution SNSPH models. Rosswog \etal
(2009) do not present a convergence study of their SNSPH models, and
find significantly less $^{56}$Ni is synthesized when they use the
finite difference code FLASH (Fryxell \etal 2009).  We are currently
using FLASH to calculate the nucleosynthesis for the three impact
parameter cases studied in this paper to investigate this apparent
discrepency and to validate (or falsify) the $^{56}$Ni yields from our
SNSPH models (Hawley \etal 2010).

\section{Discussion}
\label{sec:discussion}

Direct collisions between white dwarfs offer an unexplored mechanism
for SNIa production within the double-degenerate family of progenitor
models.  Our simulations suggest that at low impact parameters, the
collision between the most common white dwarf masses (0.6\msol) can
result in explosive nuclear burning, even though the total mass
involved is below the Chandrasekhar limit, and producing enough
$^{56}$Ni to result in a dim SNIa or be an example of a new class of
transients between Novae and SNIa.  It is likely that collisions
between more massive white dwarfs with impact parameters less than a
white dwarf radius will result in brighter events.  Although our
simulations are unable to resolve the exact physics that trigger
detonations in an open environment, these details might not be
critical for the overall nucleosynthesis in collision scenarios.

Rosswog \etal (2009) have reported that a low resolution $b=0$
collision of two 0.9\msol white dwarfs can produce enough $^{56}$Ni to
resemble the nuclear yields of a typical SNIa. As we've shown above
for two 0.6\msol white dwarfs, the nuclear yields are sensitive to the
impact parameter.  Grazing incident collisions between more massive
white dwarfs are likely to lead to similar configurations.  The
ultimate fate of these systems will, like their binary merger cousins,
depend on the total mass, thermal and neutrono cooling rates, angular
momentum transport rate, mass accretion rate onto the hot central
object, and any residual nuclear burning (Yoon \etal 2007).  Further
studies are needed to say which white dwarf collisions lead to
double-degenerate Ia supernovae or perhaps a new class of transients
between Novae and SNIa.

\section*{Acknowledgments}

This work was supported by the National Science Foundation under
grant AST 08-06720. All simulations were conducted using the Ira A. Fulton High
Performance Computing Center, Arizona State University. We thank our referee for
their suggestions and comments.

\end{document}